\def\year{2021}\relax
\documentclass[letterpaper]{article} 
\usepackage{aaai21}  
\usepackage{times}  
\usepackage{helvet} 
\usepackage{courier}  
\usepackage[hyphens]{url}  
\usepackage{graphicx} 
\usepackage{natbib}  
\usepackage{caption} 

\usepackage{multirow}
\usepackage{arydshln}

\usepackage{amsmath}

\title{Compound Word Transformer: Learning to Compose Full-Song Music \\over Dynamic Directed Hypergraphs}
\author{

    Wen-Yi Hsiao,\textsuperscript{\rm 1}
    Jen-Yu Liu,\textsuperscript{\rm 1},
    Yin-Cheng Yeh,\textsuperscript{\rm 1}
    Yi-Hsuan Yang\textsuperscript{\rm 1, \rm 2} \\
}
\affiliations{
    \textsuperscript{\rm 1}Yating Team, Taiwan AI Labs, Taiwan \\
    \textsuperscript{\rm 2}Academia Sinica, Taiwan \\
    \{wayne391,~jyliu,~yyeh,~yhyang\}@ailabs.tw \\
}

\begin{document}

\maketitle

\begin{abstract}
To apply neural sequence models such as the Transformers to music generation tasks, one has to represent a piece of music by a sequence of tokens drawn from a finite set of pre-defined vocabulary. Such a vocabulary usually involves tokens of various \emph{types}. For example, to describe a musical note, one needs separate tokens to indicate the note's pitch, duration, velocity (dynamics), and placement (onset time) along the time grid. While different types of tokens  may possess different properties, existing models usually treat them equally, in the same way as modeling words in natural languages. In this paper, we present a conceptually  different approach that explicitly takes into account the type of the tokens, such as \emph{note} types and \emph{metric} types. And, we propose a new Transformer decoder architecture that uses different feed-forward heads to model tokens of different types. With an expansion-compression trick, we convert a piece of music to a sequence of \emph{compound words} by grouping neighboring tokens, greatly reducing the length of the token sequences. We show that the resulting model can be viewed as a learner over dynamic directed hypergraphs. And, we employ it to learn to compose expressive Pop piano music of full-song length (involving up to 10K individual tokens per song), both conditionally and unconditionally. Our experiment shows that, compared to state-of-the-art models, the proposed model converges 5--10 times faster at training (i.e., within a day on a single GPU with 11 GB memory), and with comparable quality in the generated music. 
\end{abstract}

\section{Introduction}


To apply neural sequence models such as recurrent neural networks (RNNs) or Transformers \cite{vaswani2017attention} to automatic music composition (a.k.a., symbolic-domain music generation), one has to represent a  piece of music as a sequence of tokens drawn from a pre-defined \emph{vocabulary} \cite{oore2018time}. 
Unlike the case in text, such a vocabulary 
usually involves tokens of various \emph{types}.
For example, to represent a musical score, we may need tokens that describe the content of the musical notes (e.g., pitch and duration), their placement along time,
the instrument that plays each note, as well as indicators of metrical events such as the beginning of a new beat, bar (measure),  or musical phrase \cite{jazzTransformer20ismir}.
We need such a diverse set of tokens as music is 
multifaceted;
a type alone captures only a certain aspect of music (e.g., melody, harmony, rhythm, timbre) and cannot faithfully represent  a music piece.

\begin{figure}[t]
\centering
\includegraphics[width=0.99\columnwidth]{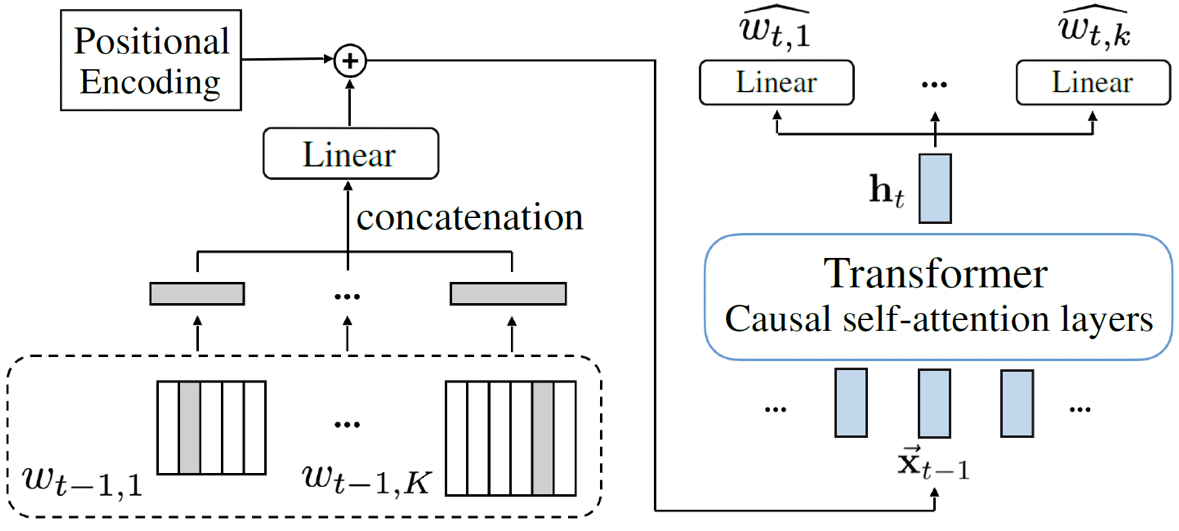}

\caption{Illustration of the main ideas of the proposed compound word Transformer: (left) \emph{compound word modeling} that combines the embeddings (colored gray) of multiple tokens $\{w_{t-1,k}\}_{k=1}^K$, one for each token type $k$, at each time step $t-1$ to form the input $\vec{\mathbf{x}}_{t-1}$  to the self-attention layers, and (right) \emph{toke type-specific feed-forward heads} that predict the list of tokens for the next time step $t$ at once at the output.}
\label{fig:architecture}
\end{figure}



As different types of (musical) tokens may have different properties, modeling the dependency of these tokens might not be the same as modeling  words in text. 
However, to our best knowledge, little work has been done to explicitly account for the heterogeneity of tokens in music. 
The tokens are mostly treated equally, in the same way as  words in text \cite{huang2018music,payne2019musenet,huang2020pop}.

\begin{table*}[t]
\centering
\small
\begin{tabular}{|l|llrrl|} 
\hline\
& Representation & Model & Attn. window & Voc. size & Data type \\ 
\hline \hline
Music Transformer \cite{huang2018music} & MIDI-like  & Transformer & 2,048 & 388 & Classical piano performance  \\ 
MuseNet \cite{payne2019musenet} & MIDI-like* & Transformer & 4,096 & N/A & Multi-track MIDI \\ 
LakhNES \cite{donahue2019lakhnes} & MIDI-like* & Transformer-XL & 512 & 630 & Multi-track MIDI  \\
TR autoencoder
\cite{choi20icml} & MIDI-like & Transformer & 2,048 & 388 & Classical piano performance    \\ 
Pop Music TR \cite{huang2020pop} & REMI & Transformer-XL & 512 & 332 & Pop piano performance    \\ 
Transformer VAE \cite{jiang20icassp} & MIDI-like & Transformer & 128 & 47 & Pop lead sheets \\ 
Guitar Transformer \cite{guitarTransformer20ismir} & REMI* & Transformer-XL & 512 & 221 & Guitar tabs \\ 
Jazz Transformer \cite{jazzTransformer20ismir} & REMI* & Transformer-XL & 512 & 451 & Jazz lead sheets \\ 
MMM \cite{ens20arxiv} & MIDI-like* & Transformer & 2,048 & $>$442  & Multi-track MIDI \\ 
\hline 
This work & CP & linear Transformer & 5,120 & 350 & Pop piano performance\\
\hline
\end{tabular}
\caption{A comparison of existing Transformer-based models and the proposed one for automatic music composition. The representations marked with * are extensions of either MIDI-like \cite{oore2018time} or REMI \cite{huang2020pop}.}
\label{tab:literature}
\end{table*}

We are therefore motivated to study in this paper whether we can improve sequence modeling of music by highlighting the role of token types.
Our first proposal is to \emph{customize the prediction heads for tokens of different types}.
Specifically, using the Transformer
as the main architecture of the underlying sequence model, we approach this by using different feed-forward heads for tokens of different types. 

Our second proposal is to group consecutive and related tokens in a token sequence into ``compound words,'' and then \emph{perform sequence modeling over} the resulting \emph{sequence of compound words}.
This is to capture the co-occurrence relationship of tokens---e.g., to generate a new musical note, we may need at least two consecutive tokens to indicate its pitch and duration; to change the tempo in the middle of a piece of music, we need a  token to indicate the target tempo value, and an co-occurring time-related token to indicate the time of the tempo change. Under the proposed compound-word modeling, the individual tokens (e.g., pitch and duration) are still predicted separately with different heads. Yet, instead of predicting them 
at different time steps, we predict multiple tokens of various types \emph{at once} in a single time step. The token embeddings of the tokens predicted at the current step are then combined and fed as the input for the next time step. Namely, the self-attention is computed over combined embeddings of individual tokens of a compound word. 

From a theoretical point of view, the proposed model can be interpreted as a learner over discrete-time \emph{dynamic directed hypergraphs} \cite{kazemi20jmlr}.
Here, a graph consists of nodes that each corresponds to a token in our vocabulary.  
A sequence of tokens can then be viewed as a sequence of edges (each connecting two nodes), or a \emph{walk}, over this graph. 
A sequence of compound words, in contrast, can be viewed as a sequence of \emph{hyperedges} (each connecting multiple nodes) \cite{feng19aaai_hypergraph}, over the same graph. 
We discuss this at greater length later in the paper.


We refer to the proposed representation as the \emph{\underline{c}om\underline{p}ound word representation}, or CP for short.
CP can be considered as an extension of  existing representations, with the following additional merits.
First, it allows for fine-grained, type-specific control over the
prediction heads.
For example, we can now use different loss functions, sampling policies, and token embedding sizes for different token types. 


Second, as a compound word represents multiple tokens at once, it requires much less time steps to generate a music piece using compound words. Namely, the sequence length of the same music piece is much shorter in CP than in existing representations. As the computational complexity of a Transformer is related to the sequence length \cite{vaswani2017attention}, this makes training and inference faster, and may facilitate learning the long-range dependency in music.\footnote{For example, we can study whether the proposed model creates music with better ``structureness,'' or  long-term repetitions  \cite{jazzTransformer20ismir,jhamtani19} in the future.}

Finally, the sequence length in CP is determined by the number of compound words in a sequence, not by the number of individual tokens per compound word. Therefore, it is possible to add new token types (by adding the corresponding feed-forward head) to increase the expressivity of the representation,  without increasing the sequence length.  
This makes it easy to extend to underlying representation, though we do not explore this potential in this work.


For performance study, we consider generating expressive Pop piano music at full-song scale in both the unconditional setting (i.e., from scratch) and conditional setting (i.e., generating the piano arrangement given the lead sheet).
This involves modeling fairly long music sequences for up to 10K individual tokens each.
We show that, with CP, we are able to train a linear Transformer decoder \cite{lineartransformer20icml} with music quality similar to that of strong 
baselines, with 
faster training and inference time. 
We provide audio examples and open source the project at a GitHub repo.\footnote{\url{https://github.com/YatingMusic/compound-word-transformer}}


\section{Related Work}

Both language and music have 
principles governing the organization of discrete structural elements (e.g., words or musical notes) into sequences \cite{patel03nature}.
As such, the Transformers, which have been firstly shown to work well for text generation \cite{child2019generating,keskar2019ctrl}, 
have been increasingly applied to 
music generation in recent years, 
by treating music pieces as sequences of discrete tokens akin to text words.
We list some related papers in Table \ref{tab:literature}.  

Table \ref{tab:literature} shows that most existing work adopt a music representation derived from either MIDI-like \cite{oore2018time} or REMI \cite{huang2020pop}, with possible addition of track- or structure-related tokens. MIDI-like and REMI differ mainly in how the advance of time is represented: the former uses [time\_shift] tokens to mark the time interval (in absolute time) 
between note-related tokens, whereas the latter assumes symbolic timing and uses [bar] and [position] tokens to place tokens on a metrical grid that uniformly divides a bar into a certain number of positions.
Neither MIDI-like nor REMI groups the tokens by token types.\footnote{Upon paper completion, we noticed an early but preliminary attempt of grouping tokens by \cite{hawthorne18workshop}.}


Existing work also differs in the length of the \emph{attention window} (see the methodology section for definition)
and vocabulary size (which is data- and task-dependent).
To our knowledge, our work represents the first one to consider Pop music modeling at full-song scale (involving 10k tokens per song), and to use the recently-proposed linear Transformer  \cite{lineartransformer20icml} as the model backbone.

\section{Methodology}


\subsection{Background}

For sequence modeling, we need a conversion function $g(\cdot)$ that converts a music piece $\mathcal{X}$ to a time-ordered sequence of symbolic elements $\mathcal{S}=g(\mathcal{X})=\{w_1, w_2, \dots, w_T\}$, where $T$ denotes the resulting sequence length.
Given a number of such sequences, we train a neural sequence model with an architecture such as the Transformer decoder to learn to generate new sequences $\mathcal{S}'$. We then use a deterministic inverse function $g^{-1}(\cdot)$ to get a new music piece from such a generated sequence, namely $\mathcal{X}'=g^{-1}(\mathcal{S}')$.
There can be different algorithms to implement the conversion function and its inverse, leading to numerous possible sequence representations of the same music piece, e.g., $\mathcal{S}_\text{MIDI-like}=g_\text{MIDI-like}(\mathcal{X})$ and $\mathcal{S}_\text{REMI}=g_\text{REMI}(\mathcal{X})$. Different conversion functions (or sequence representations) assume different vocabulary sizes $M$, so $\mathcal{S}_\text{MIDI-like}$ and $\mathcal{S}_\text{REMI}$ differ in both $T$ and $M$.

A Transformer decoder comprises a stack of \emph{self-attention} layers and a stack of \emph{feed-forward} layers. The self-attention layers operate on a fixed-length sub-sequence of $\mathcal{S}$ to learn the dependency among the  elements. The length of such a sub-sequence, a.k.a., the \emph{attention window}, denoted as $N$, is usually much smaller than $T$, as $N$ directly affects the space complexity  
of the model. For the vanilla Transformer \cite{vaswani2017attention} and its faster variant Transformer-XL \cite{dai2019transformer}, it is $\mathcal{O}(N^2M)$; for the linear Transformer \cite{lineartransformer20icml}, it is $\mathcal{O}(NM)$.


\subsection{Individual Tokens vs Compound Words}


In this paper, we refer to the elements in either $\mathcal{S}_\text{MIDI-like}$ or $\mathcal{S}_\text{REMI}$ as the \emph{individual tokens}. 
They are drawn from a pre-defined vocabulary
$\mathcal{V}=\{1,\dots,M\}$.
As mentioned in the introduction, each token is  associated with a \emph{type} defined in the type set, $\mathcal{K}=\{1,\dots,K\}$. 
We can partition $\mathcal{V}$ into $K$ subsets by token group, i.e., $\{\mathcal{V}_k\}_{k=1}^K$.

\begin{figure}[t]
\centering
\includegraphics[width=.9\linewidth]{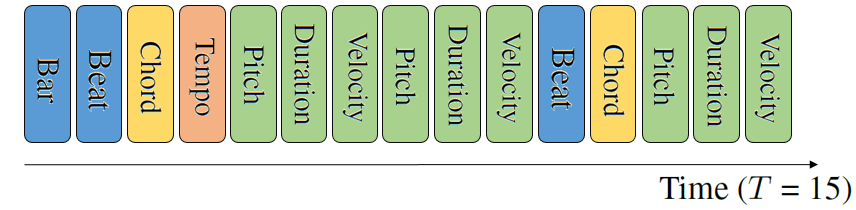}\\
(a) REMI representation  \\
\includegraphics[width=.4\linewidth]{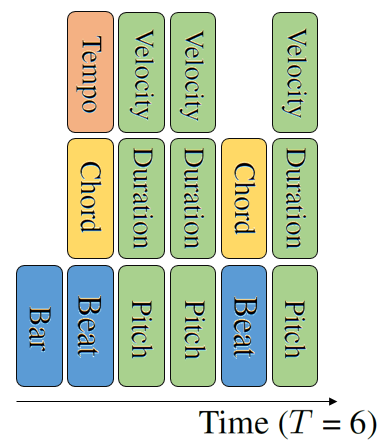} ~~~~~~~~
\includegraphics[width=.35\linewidth]{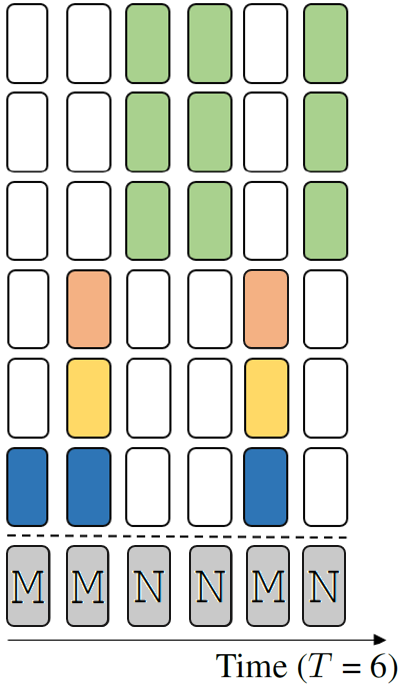}\\
(b) Tokens grouped ~~~~~~~~~~~~~
(c) Compound words
\caption{An example illustrating the conversion from a sequence of REMI tokens \cite{huang2020pop} into a (shorter) sequence of compound words. A compound word comprises a number of grouped tokens and the [ignore] tokens, which are colored white in (c), as well as a family token ($\texttt{N}$: note-related or $\texttt{M}$: metric-related). Best seen in color.}
\label{fig:repre_evolution}
\end{figure}

We propose to convert a sequence of tokens (e.g., $\mathcal{S}_\text{REMI}$) into a sequence of compound words $\mathcal{S}_\text{CP}$ with the following procedure.
First, \emph{neighboring tokens that define a musical event together are grouped into a super token}, i.e., placed on the same time step, as illustrated in Figures \ref{fig:repre_evolution}(a)--(b).
A musical event here can be a \emph{note} related one, i.e., to create a new musical note, or a \emph{metrical} related one, e.g., to mark the beginning of a new beat, or a new bar. 
For example, in REMI, a note is created by consecutive tokens of [pitch], [duration], and [velocity], which are grouped in CP.
And, a tempo or chord change in REMI takes place only at beat times, so we also group [beat], [chord] and [tempo]. Accordingly, the model has to make multiple predictions (i.e., generate multiple tokens) at each time step.

Second, we \emph{fill the missing token types per time step with ``[ignore]'' tokens}, so that at each  step there are consistently $K$ tokens to be predicted, as illustrated in Figure \ref{fig:repre_evolution}(c).
This is to make computational modeling feasible, as otherwise the shape and meaning of the target output at each time step would be uncertain. 
In other words, a \emph{compound word} is composed of a list of $K$ tokens, each drawn from the corresponding subset $\mathcal{V}_k \cup \text{[ignore]}$, that are placed on the same time step $t$.
Formally, $\mathcal{S}_\text{CP}=g_\text{CP}(\mathcal{X})=\{cp_t\}_{t=1}^{T_\text{cp}}$, in which 
$cp_t = \{w_{t,1},\cdots,w_{t,K}\}$. 
We view this conversion function $g_\text{CP}(\cdot)$ as performing an \emph{expansion-compression trick}, as the original sequence is firstly expanded to a sequence of  $KT_\text{CP}$ individual tokens, and then compressed to a sequence of $T_\text{CP}$ compound words; in general $T_\text{CP} < T_\text{REMI} < KT_\text{CP}$.

To facilitate modeling the CP, we further partition the type set $\mathcal{K}$ into $F$ \emph{families}. For example, if $\mathcal{K}$ can be partitioned into two families, the \emph{note} family $\mathcal{K}_\texttt{N}$ and \emph{metric} family $\mathcal{K}_\texttt{M}$ (marked as `\texttt{N}' and `\texttt{M}' in Figure \ref{fig:repre_evolution}(c)), we would have $\mathcal{K} = \mathcal{K}_\texttt{N} \cup \mathcal{K}_\texttt{M}$, and $\mathcal{K}_\texttt{N} \cap \mathcal{K}_\texttt{M} = \emptyset$. 
Each compound word $cp_t$ is associated with a \emph{family token} $f_t$.  For a metric-related $cp_t$, we would have $w_{t,k}=\text{[ignore]}$, for $k \in \mathcal{K}_\texttt{N}$. Similarly, for a note-related $cp_t$, $w_{t,k}=\text{[ignore]}$, for $k \in \mathcal{K}_\texttt{M}$.








\subsection{Combining Token Embeddings of Adaptive Sizes}

As input to Transformers, an element in a sequence is represented by an \emph{embedding} vector, $\mathbf{x}_t \in \mathcal{R}^d$, and then added with a positional embedding vector \cite{ke2020rethinking}. 
In CP, we propose to form an embedding vector for a compound word $cp_t$ by combining the embedding vectors $\mathbf{p}_{t,k}$ of the composing tokens $w_{t,k}$, as well as an embedding vector $\mathbf{q}_t$ associated with the family token $f_t$. Specifically, we combine the vectors by firstly concatenating them, and then linearly projecting the resulting long vector to a $d$-dimensional vector with a projection matrix $\mathbf{W}_\text{in}$. Namely,
\begin{equation}
\begin{split}
\mathbf{p}_{t,k}&= \text{Embedding}_k(w_{t,k})\,, \;  k = 1,..., K  \,, \\
\mathbf{q}_t &= \text{Embedding}_\mathcal{F}(f_t) \,, \\
\mathbf{x}_t &= \mathbf{W}_\text{in}\,[\mathbf{p}_{t,1} \oplus ... \oplus \mathbf{p}_{t,K} \oplus \mathbf{q}_t] \,, \\
\vec{\mathbf{x}}_t &= \text{Positional Encoding}(\mathbf{x}_t) \,,
\end{split}
\label{eq:1}
\end{equation}
where $\oplus$ denotes vector concatenation, and $\text{Embedding}_k(\cdot)$ and $\text{Embedding}_\mathcal{F}(\cdot)$ involve the use of lookup tables.

In essence, $\mathbf{x}_t$ can be considered as a \emph{compressive} representation of the composing tokens $w_{t,k}$ and family token $f_t$. We note the action of compressing the embeddings is reminiscent of the main idea of the Compressive Transformer \cite{rae20iclr}, which proposes to compresses past memories beyond the attention window for long-range sequence learning. Unlike it, we compress the memories \emph{within} the attention window defined over the individual tokens.

A main merit of CP is that we can customize the settings for different token types. Being inspired by the \emph{adaptive word representation} \cite{adaptiveEmbedding}, 
we use different embedding sizes $d_k$ for tokens of different types, i.e., $\mathbf{p}_{t,k} \in \mathcal{R}^{d_k}$.
We basically use larger $d_k$ for token types with larger vocabulary size $|\mathcal{V}_k|$. See Table \ref{tab:class_and_emb} for details.

\subsection{Multi-head Output Module}

A main proposal of our work is to use different feed-forward heads for tokens of different types in a Transformer. 
Specifically, we have $(K+1)$ heads in total, one for each token type $\mathcal{V}_k$ and an additional one for the token family $\mathcal{F}$.

Instead of working on the $K+1$ heads at the same time, we devise a \emph{two-stage} setting that predicts the family token first, and then the remaining tokens given the family token. 
Specifically, at the $t$-th time step, the feed-forward procedure can be summarized as:
\begin{equation}
\begin{split}
\mathbf{h}_t &= \text{Self-attn} \left(\vec{\mathbf{x}}_{t-1} \right)  \,, \\
\widehat{f_t} &= \text{Sample}_\mathcal{F} \left( \text{softmax}(\mathbf{W}_\mathcal{F}\mathbf{h}_t)  \right) \,,  ~~~~~\\
\mathbf{h}_t^\text{out} &= \mathbf{W}_\text{out}\,[\mathbf{h}_t \oplus \text{Embedding}_\mathcal{F}(\widehat{f_t}) ] \,, ~~~~ \\
\widehat{w_{t,k}} &= \text{Sample}_k \left( \text{softmax}(\mathbf{W}_k\mathbf{h}_t^\text{out}) \right) \,, \;  k = 1,..., K  \,,
\end{split}
\label{eq:2}
\end{equation}
where $\mathbf{W}_\mathcal{F}$ and $\{\mathbf{W}_k\}_{k=1}^K$
are the $K+1$ feed-forward heads, 
$\text{Self-attn}(\cdot)$ the causal self-attention layers, and $\text{Sample}(\cdot)$ a sampling function.
We empirically find that this two-stage setting makes it easier for the model to predict $w_{t,k}=\text{[ignore]}$, for $k$ not in the target family $\mathcal{K}_{\widehat{f_t}}$. 

Figure \ref{fig:architecture} illustrates  Eqs. (\ref{eq:1})--(\ref{eq:2}) in work, omitting the first-stage part at the output for $\widehat{f_t}$ due to space limit.




\subsection{Adaptive Sampling Policy}
At inference time, we use stochastic temperature-controlled  sampling \cite{holtzman20iclr} to avoid degeneration and to increase diversity. 
With CP, we employ different sampling policies $\text{Sample}_k(\cdot)$ for different token types; see Table \ref{tab:class_and_emb}.


\section{Graph Interpretation}


We discuss the proposed model from a graph-theoretical point of view below.
Given a vocabulary of tokens, we can construct a fully-connected \emph{static graph} $\mathcal{G}=(\mathcal{V},\mathcal{E})$ \cite{kivela14jcn} comprising nodes $\mathcal{V}=\{1,\dots,M\}$ and edges $\mathcal{E} = \mathcal{V} \times \mathcal{V}$.  Each node corresponds to an individual token in our vocabulary.  
This way, a token sequence $\mathcal{S}_\text{MIDI-like}$ or $\mathcal{S}_\text{REMI}$ can be viewed as a sequence of edges (each connecting two nodes), or a \emph{walk}, over this graph. 

In CP, the vocabulary (and accordingly the graph) is augmented with a set  of \emph{special tokens}, denoted as $\mathcal{V}^*$, that includes for example type-specific [ignore] tokens and family tokens. 
And, a compound word consists of $K+1$ nodes, one from each of the $K$ types and an additional one from the set of family tokens. A sequence of compound words, namely $\mathcal{S}_\text{CP}$, therefore, involves transitions from $K+1$ nodes to another $K+1$ nodes per time step.
Such a transition can be viewed as a directed  \emph{hyperedge} \cite{feng19aaai_hypergraph,jiang19ijcai}, 
that connects at once $K+1$ source nodes (e.g., $cp_{t-1}$) to $K+1$ target nodes ($cp_t$).
It is directed because the order of the nodes matters (i.e., from $t-1$ to $t$).

A sequence of compound words also forms a
\emph{dynamic directed hypergraph}  \cite{kazemi20jmlr}:  $\{\mathcal{G}_1, \mathcal{G}_2, \dots, \mathcal{G}_T\}$, where $\mathcal{G}_t = (\mathcal{V},\mathcal{E}_t)$. Starting from an empty graph with no edges, at each time step $t>1$ we add a new directed hyperedge, labeled with the time step $t$, connecting in total $2K+2$ nodes.
In practice, we have a [BOS] token (beginning of sequence) and [EOS] token (end of sequence), so the hyperedge at $t=1$ and $t=T$ connects to only $K+2$ nodes.

A neural model for graphs, or a
\emph{graph neural network} (GNN), can be regarded as an encoder-decoder pair \cite{kazemi20jmlr,rossi20arxiv}, where an \emph{encoder} is a function that maps from a graph $\mathcal{G}$ to node embeddings $\mathbf{z}_i, i=1 \dots M$, and a \emph{decoder} takes as input one ore more node embeddings and makes a prediction based on these, e.g., node classification or edge prediction.
The proposed CP Transformer can therefore be regarded as a learner over dynamic directed hypergraphs, as at each time step $t$ it manages to predict the next hyperedge to be added (i.e., $\widehat{w_{t,k}}$ and $\widehat{f_t}$) based on the node embeddings updated from $\mathcal{G}_{<t}=\{\mathcal{G}_1, \mathcal{G}_2, \dots, \mathcal{G}_{t-1}\}$, or the collection of input embeddings $\mathbf{x}_{<t}=\{\mathbf{x}_1, \mathbf{x}_2, \dots, \mathbf{x}_{t-1}\}$ marked with positional embeddings (i.e., edge labels on the directed hyperedges).

We note that, while we introduce the proposed methods in the context of music modeling, the idea of compound words is generic and may be applicable to sequences seen in other data domains, when multiple tokens (i.e., a hyperedge) are needed to represent a single event, entity, or object.

\section{Implementation}

To test the effectiveness of the proposed methods, we implement a CP Transformer that learns to generate Pop piano music with human performance characteristics such as expressive variations in velocity 
(i.e., the force with which a note is played, which is related to loudness) 
and tempo \cite{oore2018time,alex2019music}. We consider Pop piano 
for its richness and expressivity, 
and for offering a direct performance comparison with the Pop Music Transformer \cite{huang2020pop} (see Table \ref{tab:literature}).


Specifically, we consider both the \textbf{conditional} and \textbf{unconditional} generation tasks. 
In the former, a \emph{lead sheet} (i.e., a melody line and an accompanying sequence of chord labels) is given, and the model has to generate a piano performance according to that. In the latter, the model  generates a piano performance  of full-song length from scratch freely. 

We intend to compare CP with REMI 
in our evaluation. We provide the implementation details below.


\begin{table}[t]
\centering
\small
\begin{tabular}{|ll|cr|} 
\hline\
\multirow{2}{*}{Task} &\multirow{2}{*}{Repre.} &\multicolumn{2}{c|}{\#words ($T$)} \\
 & & mean ($\pm$ std)~~~~ & max \\
 \hline  \hline
\multirow{2}{*}{Conditional} &REMI  &6,432 ($\pm$ 1,689)  & 10,240  \\ 
            &CP     &3,142 ($\pm$ 821)~~  & 5,120  \\
\hline 
\multirow{2}{*}{Unconditional} &REMI   & 4,873 ($\pm$ 1,311)  & 7,680 \\
             &CP    & 2,053  ($\pm$ 580)~~ &  3,584 \\
\hline
\end{tabular}
\caption{Statistics of the number (\#) of words (i.e., tokens in REMI; compound words in CP) per song in the training set.}
\label{tab:training_data_stats}
\end{table}

\subsection{Dataset}
We collect the audio files of 1,748 pieces of Pop piano from the Internet. 
The average length of the songs is about 4 minutes, and we have about 108 hours in total.  All the songs are in 4/4 time signature (four beats per bar).
We convert each song 
(an audio) into a symbolic sequence  
as follows.

\begin{itemize}
\item \textbf{Transcription}: We use the state-of-the-art RNN model for automatic piano transcription, ``Onset and Frames'' \cite{hawthorne2018onsets}, to estimate the pitch, onset and offset time, and velocity of the musical notes from audio.

\item \textbf{Synchronization}: To get symbolic timing from the original wall clock time, we use the RNN-based model available in the Python package \texttt{madmom} \cite{madmom} to estimate the downbeat and the beat positions, which represent the state-of-the-art for the task.
Then, we interpolate 480 ticks between two adjacent beats, and map the absolute time into its according tick. By doing so, we can keep tiny offset. Lastly, we infer the tempo changes from the time interval between adjacent beats. 

\item \textbf{Quantization}: We quantize the tempo, velocity, duration and the beat positions to reduce the size of the vocabulary. For example, we set the 16-th note as our basic time unit. See Table \ref{tab:class_and_emb} for the number of tokens per type. 

\item \textbf{Analysis}: For the conditional generation task, we estimate the melody notes and chord symbols from the  transcription result to form the lead sheets. Specifically, we develop an in-house rule-based chord recognition algorithm\footnote{\url{https://github.com/joshuachang2311/chorder}}
to recognize 12 roots and 7 chord qualities. We use the ``Skyline algorithm'' \cite{uitdenbogerd99mm} 
to extract the melodies. And, as a lead sheet is usually of coarser time resolution, we quantize the chord symbols and melody notes to the 4-th notes (i.e., beat times).
\end{itemize}
We randomly hold out 50 songs for testing, and use the remaining for training the Transformers.  

\begin{table}[t]
\centering
\small

\begin{tabular}{|l|l|r|r|cc|} 
\hline
\multirow{2}{*}{Repre.} & \multirow{2}{*}{Token type} & Voc. size  &Embed. & \multicolumn{2}{|c|}{$\text{Sample}_k(\cdot)$} \\
 & & $|\mathcal{V}_k|$ &size ($d_k$) &$\tau$ &$\rho$\\
\hline\hline
\multirow{9}{*}{CP}   & [track] &2 (+1)   &3   &1.0  & 0.90 \\
    \cdashline{2-6}[1pt/1pt]
     & [tempo]  &58 (+2)   &128 &1.2 & 0.90  \\
     & [position/bar]  &17 (+1)   &64  &1.2 & 1.00 \\
     & [chord]  &133 (+2)   &256 &1.0  &  0.99 \\
     \cdashline{2-6}[1pt/1pt]
     & [pitch] &86 (+1)   &512 &1.0  & 0.90 \\
     & [duration] &17 (+1)   &128 &2.0   & 0.90 \\
     & [velocity] &24 (+1)   &128 &5.0   & 1.00 \\
     \cdashline{2-6}[1pt/1pt]
     & [family]  &4  ~~~~~~~   &32  &1.0  &0.90 \\
     \cdashline{2-6}[1pt/1pt]
     &total  &341  (+9) & ---& --- & ---\\
\hline\hline
REMI &total &338 ~~~~~~~    &512 &1.2 &0.90 \\
\hline
\end{tabular}
\caption{Details of the CP representation in our implementation, including 
that of the
sampling policy ($\tau$-tempered  top-$\rho$ sampling). For the vocabulary size, the values in the parentheses denote the number of special tokens such as [ignore].}
\label{tab:class_and_emb}
\end{table}



\subsection{Vocabulary}

To represent the content of a piano performance, the basic setting employs tokens of six types: three note-related types [pitch], [duration],  [velocity], and three metric-related types [position/bar], [tempo], [chord].
The specific vocabulary is task-dependent and is introduced below.

\emph{Conditional generation}---We additionally use [track] tokens to mark whether it is the \emph{lead sheet} track (i.e., the condition) or the \emph{piano} track (the track to be generated). While the piano track (i.e., the sub-sequence after the [track=piano] token) involves all the six types of tokens mentioned above, the lead sheet track only involves the use of composition-related tokens [position/bar], [chord], [pitch], [duration], not performance-related tokens [velocity], [tempo].
In CP, we have three family tokens, [family=track], [family=note], [family=metric]. 
Moreover, we have type-specific [ignore] tokens and an additional [conti] token for the beat positions having 
no tempo or chord changes. 





\emph{Unconditional generation}---This task only concerns with the piano track so we do not need the [track] tokens. But, as it concerns with full-song generation, we add an [EOS] token to signify the end of a sequence. 
We view it as a family token, so there are three possible family tokens here: [family=EOS], [family=note], [family=metric]. 




Details of the adopted representations are shown in Tables \ref{tab:training_data_stats} and \ref{tab:class_and_emb}. 
Table \ref{tab:training_data_stats} compares the sequence length $T$ of REMI and CP. We can see that $\mathcal{S}_\text{CP}$ is much shorter than $\mathcal{S}_\text{REMI}$, especially under the conditional task.\footnote{We set an upper limit of the number of elements per sequence (e.g., 10,240 tokens in REMI) and remove overly long songs, which amounts to removing 25--88 songs from the training set depending on the task and the adopted representation.}
Table \ref{tab:class_and_emb} displays the size of each vocabulary subset $\mathcal{V}_k$. 
We see that CP and REMI have similar total vocabulary size $M$. REMI does not use the family tokens (except for [EOS]) and special tokens.



\begin{table*}[t]
\centering
\small
\begin{tabular}{|llrr|rr|rr|} \hline
 \multirow{2}{*}{Task} & \multirow{2}{*}{Representation $+$ model@loss}  & Training &GPU   & \multicolumn{2}{c|}{Inference (/song)} & \multicolumn{2}{c|}{Matchness} \\
 & &time &memory & time (sec) & tokens (\#) & melody & chord \\
\hline\hline
\multirow{6}{*}{Conditional}
&Training data &---&---&---&---&0.755 &0.838 \\
&Training data (randomized) &---&---&---&---&0.049 &0.239 \\
\cdashline{2-8}[1pt/1pt]
&REMI $+$ XL@0.44  &3 days &4 GB   &88.4 & 4,782 & 0.872 & 0.785\\
&REMI $+$ XL@0.27   &7 days &4 GB &91.5 & 4,890   & 0.866 & 0.800 \\
\cdashline{2-8}[1pt/1pt]
&REMI $+$ linear@0.50 & 3 days &17 GB &48.9 & 4,327 & 0.779 & 0.709 \\
&CP~~~~~ $+$ linear@0.27 & 0.6 days  &10 GB &29.2 & 18,200 & 0.829 & 0.733\\
\hline
\multirow{2}{*}{Unconditional}
&REMI $+$ XL@0.50      & 3 days & 4 GB &139.9 &7,680 &---&--- \\
&CP~~~~~ $+$ linear@0.25 & 1.3 days &9.5 GB &19.8 & 9,546 &---&--- \\
\hline
\end{tabular}
\caption{Quantitative  evaluation result 
of different models. REMI$+$XL represents a re-implementation of the state-of-the-art Pop Music Transformer \cite{huang2020pop}, while CP$+$linear stands for the proposed CP Transformer.}
\label{tab:efficiency_table}
\end{table*}



\subsection{Model Settings}


For the backbone architecture of  our model, we employ the linear Transformer \cite{lineartransformer20icml},\footnote{\url{https://github.com/idiap/fast-transformers}} as its complexity is a linear function of the length of the attention window $N$.  Moreover, we set $N$ equal to the sequence length $T$ for our model. That is, \emph{no segmentation} over the training sequences is done, and thereby \emph{all the tokens in a sequence can be accessed} by our model under causal masking, without using tricks such as memory caching \cite{dai2019transformer} or memory compression \cite{rae20iclr}. We  refer to our model as \textbf{CP$+$linear} in what follows.

For the \textbf{baselines}, we employ the Pop Music Transformer \cite{huang2020pop}, which is open-source and stands for a state-of-the-art for unconditional music composition.\footnote{\url{https://github.com/YatingMusic/remi}} This \textbf{REMI$+$XL} model adopts the REMI representation and uses Transformer-XL \cite{dai2019transformer} as the model backbone. As its complexity grows quadratically with $N$, we set $N=512$, following \cite{huang2020pop}.

Moreover, we consider one more baseline that replaces  Transformer-XL by linear Transformer, using also $N=T$, to offer a sensible performance comparison between CP and REMI. We refer to this variant as \textbf{REMI$+$linear}. 

We use 12 self-attention layers each with 8 attention heads for all the models for fair comparison. The model hidden size and inner layer of the feed-forward part are set to 512 and 2,048, respectively. 
For the token embedding size $d$, we fix it to 512 for REMI, following \cite{huang2020pop}. For CP, we set it adaptively based on the vocabulary size 
of each token type, as shown in Table \ref{tab:class_and_emb}. 
For sampling, we employ the ``nucleus sampling'' \cite{holtzman20iclr}, a stochastic method that samples from the smallest subset of tokens whose cumulative probability mass  exceeds a threshold $\rho \in [0,1]$. Before sampling, we reshape the probability distribution of the tokens (e.g., softmax($\mathbf{W}_k\mathbf{h}_t^\text{out}$)) through ``temperature'' \cite{temperature}, with the temperature parameter $\tau>0$. As Table \ref{tab:class_and_emb} also shows, we use different $\rho$ and $\tau$ for different token types. For example, we use a large $\tau$ 
to encourage diverse velocity values.



The conditional generation task can be approached with a sequence-to-sequence model, since we have paired data of lead sheets and piano performances (i.e., the former is extracted automatically from the latter). 
Instead of adding a Transformer encoder  (as done in \cite{choi20icml})
to realize this, we use the encoder-free ``Prefix LM'' method of the Google's ``T5'' model \cite{raffel20jmlr_t5}, and run a single Transformer 
over an \emph{interleaved} sequence of lead sheets and piano performances. Specifically, a sequence of lead sheet and the corresponding target sequence of piano performance are integrated into one sequence bar after bar. That is, the integrated sequence would have the form of \{$\dots$, [bar], [track=leadsheet], (content of the lead sheet for a bar), [track=piano], (content of the piano for the same bar), [bar], (content of the two tracks of the next bar) $\dots$\}. This makes it easy to learn the dependency of the two tracks, and to impose the pre-given lead sheet at inference time.


\section{Quantitative Evaluation}

The experiments hereafter are conducted in the interest of a resource-constrained scenario, assuming that we only have a single GPU with 11 GB memory and are only willing to train a model for 3 days. We conjecture that this makes sense for most middle-size academic labs worldwide.
Yet, to have an idea of the model performance when more resources are available, we include to the evaluation of the conditional task two settings exceeding such a specification.


We firstly compare the efficiency of the models in terms of training time, inference time, and GPU memory usage, under the conditional setting. The average result over the 50 held-out test songs is shown in Table \ref{tab:efficiency_table}.

\textbf{GPU memory usage}.  Table \ref{tab:efficiency_table} shows that both CP$+$linear and REMI$+$XL require $<$11\,GB GPU memory for training. Accordingly, in our implementation, we train them (separately) on an NVIDIA RTX 2080 Ti GPU (with 11GB memory). In contrast, REMI$+$linear requires 17\,GB GPU memory, so we train it on a TITAN GPU with 24\,GB memory.

\textbf{Training time}. We see that REMI-based models require much longer clock time to reach a low training loss. While it takes nearly 7 days for REMI$+$XL to reduce the negative log-likelihood (NLL) of the training data to 0.27, it takes only 0.6 days for CP$+$linear to reach the same NLL. Such a training efficiency is desirable (especially given that it is on a single 2080 Ti GPU), as it makes further extensions and modifications of the model easy and affordable.

\textbf{Inference time}. CP$+$linear is remarkably fast, taking on average  $<$30 seconds to complete the conditional generation of a song. 
As a song in our dataset is about 4 minutes, this is much faster than real time. In contrast, REMI$+$XL and REMI$+$linear are about 3x and 1.7x  slower, respectively. CP$+$linear is fast for it generates in total 8 individual tokens (of different types) at once  each time step. 


Table \ref{tab:efficiency_table} also compares the efficiency of REMI$+$XL and CP$+$linear under the unconditional setting, for which we generate also 50 songs (from scratch) and report the average inference time. We see that CP$+$linear is even faster here, requiring only $<$20 seconds to create a new  song at full-song length. In contrast, REMI$+$XL is on average 7x  slower.


\begin{table}[t]
\centering
\small
\begin{tabular}{|l|cccc|c|} \hline
Repre. $+$ model@loss  &\textbf{F} &\textbf{R} &\textbf{H} &\textbf{C} &\textbf{O} \\
\hline\hline
REMI $+$ XL@0.44        &4.05         &3.12 &3.38 &3.55 &3.31 \\
REMI $+$ XL@0.27        &\textbf{4.29} &\textbf{3.14} &\textbf{3.70} &\textbf{3.64} &\textbf{3.35} \\
REMI $+$ linear@0.50    &4.03          &3.09 &3.48 &3.46 &3.29 \\
CP~~~~~ $+$ linear@0.27 &4.09          &3.13 &3.50 &3.31 &3.08 \\
\hline
\end{tabular}

\vspace{0.15cm}
\normalsize{(a) Conditional generation}
\vspace{0.3cm}
\small

\begin{tabular}{|l|ccc|c|} \hline
Repre. $+$ model@loss   &\textbf{R} &\textbf{H} &\textbf{S} &\textbf{O} \\
\hline\hline

REMI $+$ XL@0.50         &3.11 &3.46 &2.91 &3.03 \\
CP~~~~~ $+$ linear@0.22  &\textbf{3.33} &\textbf{3.68} &\textbf{3.11} &\textbf{3.34} \\
\hline
\end{tabular}

\vspace{0.15cm}
\normalsize{(b) Unconditional generation}

\caption{Result of subjective evaluation (\textbf{F}idelity, \textbf{R}ichness, \textbf{H}umanness, \textbf{C}orrectness, \textbf{S}tructureness, \textbf{O}verall). }
\label{tab:user_study}
\end{table}

Next, we compare the performance of the models in terms of two objective metrics, also under the conditional setting.  
As the goal is to generate a song given a lead sheet, we can measure whether the generated song has a melody line and chord progression similar to that in the given condition, and take that as a figure of merit. (In contrast, proper objective evaluation of unconditional generation models remains an open issue \cite{yang20evaluation,dong20ismir_muspy,jazzTransformer20ismir}.)
Specifically, we consider:
\begin{itemize}
    \item \textbf{Melody matchness}. We represent the lead sheet and the correspondingly generated piano both in the REMI format and compute the bar-wise \emph{longest common sub-sequence} (LCS) of the two resulting sequences $\mathcal{S}_\text{REMI}^\text{LS}$ and $\widehat{\mathcal{S}}_\text{REMI}^\text{piano}$. When two notes (each from the two sequences) have the same pitch and close onset time (within the 8-th note), we consider that as a match. 
    We divide the length of the LCS by the number of [pitch] tokens in $\mathcal{S}_\text{REMI}^\text{LS}$ (i.e., the number of target melody notes) of that bar, and take the average value of such a ratio across all the bars of a song as a simple measure of melody matchness.
    \item \textbf{Chord matchness}. The \emph{chroma vector} \cite{fujishima99} represents a short-time fragment of music by the distribution of energy across the 12 pitch classes (\texttt{C}, \texttt{C\#}, etc) and offers a simple way to evaluate the harmonic similarity between two fragments. 
    We calculate the segment-wise cosine similarity between the chroma vector representing each chord label of a lead sheet (which would be binary-valued) and the chroma vector of the correspondingly generated piano segment (normalized by the maximum value so it is $\in [0,1]^{12}$), and treat the average value across time as a measure of chord matchenss.
\end{itemize}


Table \ref{tab:efficiency_table} shows that the evaluated models all have matchness close to that of the training set, and much higher than that of the random baseline (i.e., the average matchness between a lead sheet and a random song from the test set). 
This suggests, while CP$+$linear is  easier and faster to train than REMI$+$XL, they may generate music of similar quality. We further investigate this through a user study, which directly assesses the perceptual quality of the generated music.



\begin{figure}[t]
\centering
\includegraphics[width=\columnwidth]{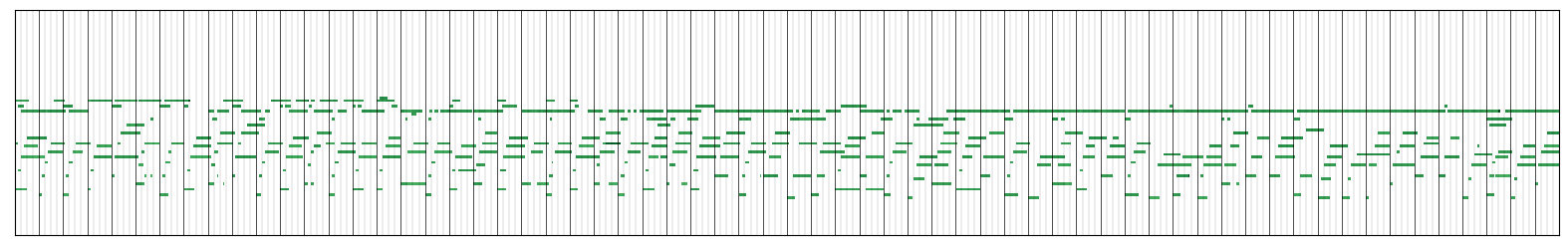}
\vspace{0.05cm}
\normalsize{(a) REMI$+$XL}
\includegraphics[width=\columnwidth]{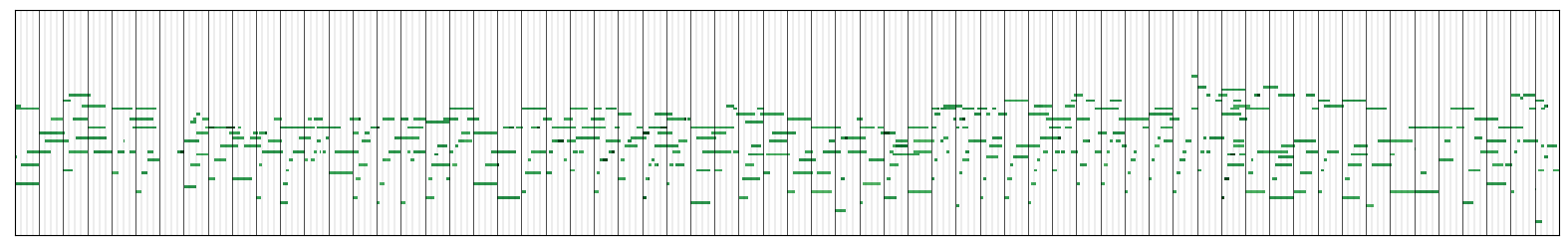}
\normalsize{(b) CP$+$linear}
\\
\caption{Piano-rolls of middle 64 bars of random generated pieces of two models in the unconditional setting. We see richer and diverse content in the result of CP$+$linear.
}
\label{fig:fs_song_example}
\end{figure}

\section{Qualitative Evaluation}


We devise an online questionnaire that solicits anonymous response to the  music generated by different models for both the conditional and unconditional settings. 
For the former, we present excerpts of 32 bars taking from one-third location of the music. For the latter, we present the full songs (i.e., when an [EOS] token is generated).\footnote{It turns out that the REMI$+$XL model seldom generates [EOS] tokens even when the music is already quite long (e.g., 8 minutes), so we stop it each time when it has generated 7,680 tokens.} 
Our intention is to investigate whether CP$+$linear and REMI$+$XL indeed generate music of similar perceptual qualities. 

The generated music is rendered into audio with a piano synthesizer using a free, non-professional grade sound font.
Each batch comprises the result of the evaluated models in random order. 
A subject has to rate the music for three random batches for each setting separately, in terms of the following aspects on a five-point Likert scale. 
1)\,\textbf{Fidelity}: is the conditionally generated piece similar to the reference, from which the condition lead sheet was taken from? 2)\,\textbf{Richness}: diversity and interestingness.
3)\,\textbf{Humanness}: does the piece 
sound like expressive human performances?
4)\,\textbf{Correctness}: perceived absence of composing or playing mistakes.
5)\,\textbf{Structureness}: whether there are structural patterns such as repeating themes or development of musical ideas. 
6)\,\textbf{Overall}. 
As the music can be long, the questionnaire may take around 30 mins to complete.


Table \ref{tab:user_study} shows the average result from 18 subjects. 
We see that REMI$+$XL performs the best in the conditional setting, yet with only moderate performance gap between the models.\footnote{In the conditional setting, the global structure of the song to be generated is fairly outlined in the given condition (i.e., the melody). Thus, it seems sufficient for models to learn from short segments.}
In contrast, CP$+$linear performs (slightly) better consistently across the four metrics in the unconditional setting, suggesting it a powerful alternative to REMI$+$XL.




\section{Conclusion}

In this paper, we have presented a new variant of the Transformer that processes multiple consecutive tokens at once at a time step. Each individual token is associated with a token type, which is exploited by the model to customize its input and output modules. The proposed model achieves sequence compression by integrating the embeddings of the tokens, which can be seen as forming a hyperedge over a dynamic graph. We show that the new Transformer works remarkably well for modeling  music, creating full-song piano of comparable perceived quality with a competing Transformer-XL based  model in much shorter training and inference time.

\section{Ethics Statement}

Research on automatic music generation may infringe copyright laws and may raise concerns regarding the role of human musicians in the future.
Cares have to be given regarding the fair use of existing musical material for model training, and the potential concern of ``deepfaking'' an existing artist's style in computer-generated music.

\section{Acknowledgement}

We are grateful to our interns at the Taiwan AI Labs, Joshua Chang for developing the symbolic-domain chord recognition algorithm, and Yu-Hua Chen and Hsiao-Tzu Hung for helping organize the PyTorch code. We also thank the anonymous reviewers for their valuable comments.


\bibliography{ref}

\begin{thebibliography}{36}
\providecommand{\natexlab}[1]{#1}
\providecommand{\url}[1]{\texttt{#1}}
\providecommand{\urlprefix}{URL }
\expandafter\ifx\csname urlstyle\endcsname\relax
  \providecommand{\doi}[1]{doi:\discretionary{}{}{}#1}\else
  \providecommand{\doi}{doi:\discretionary{}{}{}\begingroup
  \urlstyle{rm}\Url}\fi

\bibitem[{Ackley, Hinton, and Sejnowski(1985)}]{temperature}
Ackley, D.~H.; Hinton, G.~E.; and Sejnowski, T.~J. 1985.
\newblock A learning algorithm for Boltzmann machines.
\newblock \emph{Cognitive Science} 9(1): 147--169.

\bibitem[{Baevski and Auli(2018)}]{adaptiveEmbedding}
Baevski, A.; and Auli, M. 2018.
\newblock Adaptive input representations for neural language modeling.
\newblock \emph{arXiv preprint arXiv:1809.10853} .

\bibitem[{B\"{o}ck et~al.(2016)B\"{o}ck, Korzeniowski, Schl\"{u}ter, Krebs, and
  Widmer}]{madmom}
B\"{o}ck, S.; Korzeniowski, F.; Schl\"{u}ter, J.; Krebs, F.; and Widmer, G.
  2016.
\newblock Madmom: A new Python audio and music signal processing library.
\newblock In \emph{Proc. ACM Multimedia}, 1174–1178.

\bibitem[{Chen et~al.(2020)Chen, Huang, Hsiao, and
  Yang}]{guitarTransformer20ismir}
Chen, Y.-H.; Huang, Y.-S.; Hsiao, W.-Y.; and Yang, Y.-H. 2020.
\newblock Automatic composition of guitar tabs by Transformers and groove
  modeling.
\newblock In \emph{Proc. Int. Soc. Music Information Retrieval Conf.}

\bibitem[{Child et~al.(2019)Child, Gray, Radford, and
  Sutskever}]{child2019generating}
Child, R.; Gray, S.; Radford, A.; and Sutskever, I. 2019.
\newblock Generating long sequences with sparse Transformers.
\newblock \emph{arXiv preprint arXiv:1904.10509} .

\bibitem[{Choi et~al.(2020)Choi, Hawthorne, Simon, Dinculescu, and
  Engel}]{choi20icml}
Choi, K.; Hawthorne, C.; Simon, I.; Dinculescu, M.; and Engel, J. 2020.
\newblock Encoding musical style with transformer autoencoders.
\newblock In \emph{Proc. Int. Conf. Machine Learning}.

\bibitem[{Dai et~al.(2019)Dai, Yang, Yang, Carbonell, Le, and
  Salakhutdinov}]{dai2019transformer}
Dai, Z.; Yang, Z.; Yang, Y.; Carbonell, J.; Le, Q.; and Salakhutdinov, R. 2019.
\newblock Transformer-{XL}: Attentive language models beyond a fixed-Length
  context.
\newblock In \emph{Proc. Annual Meeting of the Association for Computational
  Linguistics}, 2978--2988.

\bibitem[{Donahue et~al.(2019)Donahue, Mao, Li, Cottrell, and
  McAuley}]{donahue2019lakhnes}
Donahue, C.; Mao, H.~H.; Li, Y.~E.; Cottrell, G.~W.; and McAuley, J. 2019.
\newblock {LakhNES}: Improving multi-instrumental music generation with
  cross-domain pre-training.
\newblock In \emph{Proc. Int. Soc. Music Information Retrieval Conf.},
  685--692.

\bibitem[{Dong et~al.(2020)Dong, Chen, McAuley, and
  Berg-Kirkpatrick}]{dong20ismir_muspy}
Dong, H.-W.; Chen, K.; McAuley, J.; and Berg-Kirkpatrick, T. 2020.
\newblock {MusPy}: A toolkit for symbolic music generation.
\newblock In \emph{Proc. Int. Soc. Music Information Retrieval Conf.}

\bibitem[{Ens and Pasquier(2020)}]{ens20arxiv}
Ens, J.; and Pasquier, P. 2020.
\newblock {MMM}- Exploring conditional multi-track music generation with the
  Transformer.
\newblock \emph{arXiv preprint arXiv:2008.06048} .

\bibitem[{Feng et~al.(2019)Feng, You, Zhang, Ji, and
  Gao}]{feng19aaai_hypergraph}
Feng, Y.; You, H.; Zhang, Z.; Ji, R.; and Gao, Y. 2019.
\newblock Hypergraph neural networks.
\newblock In \emph{Proc. AAAI}, 3558--3565.

\bibitem[{Fujishima(1999)}]{fujishima99}
Fujishima, T. 1999.
\newblock Realtime chord recognition of musical sound: A system using common
  {L}isp.
\newblock In \emph{Proc. International Computer Music Conf.}, 464--467.

\bibitem[{Hawthorne et~al.(2018{\natexlab{a}})Hawthorne, Elsen, Song, Roberts,
  Simon, Raffel, Engel, Oore, and Eck}]{hawthorne2018onsets}
Hawthorne, C.; Elsen, E.; Song, J.; Roberts, A.; Simon, I.; Raffel, C.; Engel,
  J.; Oore, S.; and Eck, D. 2018{\natexlab{a}}.
\newblock {Onsets and Frames}: Dual-objective piano transcription.
\newblock In \emph{Proc. Int. Soc. Music Information Retrieval Conf.}, 50--57.

\bibitem[{Hawthorne et~al.(2018{\natexlab{b}})Hawthorne, Huang, Ippolito, and
  Eck}]{hawthorne18workshop}
Hawthorne, C.; Huang, A.; Ippolito, D.; and Eck, D. 2018{\natexlab{b}}.
\newblock Transformer-{NADE} for piano performances.
\newblock In \emph{Proc. Machine Learning for Creativity and Design Workshop}.

\bibitem[{Holtzman et~al.(2020)Holtzman, Buys, Du, Forbes, and
  Choi}]{holtzman20iclr}
Holtzman, A.; Buys, J.; Du, L.; Forbes, M.; and Choi, Y. 2020.
\newblock The curious case of neural text degeneration.
\newblock In \emph{Proc. Int. Conf. Learning Representations}.

\bibitem[{Huang et~al.(2019)Huang, Vaswani, Uszkoreit, Simon, Hawthorne,
  Shazeer, Dai, Hoffman, Dinculescu, and Eck}]{huang2018music}
Huang, C.-Z.~A.; Vaswani, A.; Uszkoreit, J.; Simon, I.; Hawthorne, C.; Shazeer,
  N.; Dai, A.~M.; Hoffman, M.~D.; Dinculescu, M.; and Eck, D. 2019.
\newblock {Music Transformer}: Generating music with long-term structure.
\newblock In \emph{Proc. Int. Conf. Learning Representations}.

\bibitem[{Huang and Yang(2020)}]{huang2020pop}
Huang, Y.-S.; and Yang, Y.-H. 2020.
\newblock {Pop Music Transformer}: Beat-based modeling and generation of
  expressive {Pop} piano compositions.
\newblock In \emph{Proc. ACM Multimedia}.

\bibitem[{Jhamtani and Berg-Kirkpatrick(2019)}]{jhamtani19}
Jhamtani, H.; and Berg-Kirkpatrick, T. 2019.
\newblock Modeling Self-Repetition in Music Generation using Generative
  Adversarial Networks.
\newblock In \emph{Proc. Machine Learning for Music Discovery Workshop}.

\bibitem[{Jiang et~al.(2019)Jiang, Wei, Feng, Cao, and Gao}]{jiang19ijcai}
Jiang, J.; Wei, Y.; Feng, Y.; Cao, J.; and Gao, Y. 2019.
\newblock Dynamic hypergraph neural networks.
\newblock In \emph{Proc. IJCAI}, 2635--2641.

\bibitem[{{Jiang} et~al.(2020){Jiang}, {Xia}, {Carlton}, {Anderson}, and
  {Miyakawa}}]{jiang20icassp}
{Jiang}, J.; {Xia}, G.~G.; {Carlton}, D.~B.; {Anderson}, C.~N.; and {Miyakawa},
  R.~H. 2020.
\newblock {Transformer VAE}: A hierarchical model for structure-aware and
  interpretable music representation learning.
\newblock In \emph{Proc. Int. Conf. Acoustics, Speech and Signal Processing},
  516--520.

\bibitem[{Katharopoulos et~al.(2020)Katharopoulos, Vyas, Pappas, and
  Fleuret}]{lineartransformer20icml}
Katharopoulos, A.; Vyas, A.; Pappas, N.; and Fleuret, F. 2020.
\newblock Transformers are {RNNs}: Fast autoregressive Transformers with linear
  attention.
\newblock In \emph{Proc. Int. Conf. Machine Learning}.

\bibitem[{Kazemi et~al.(2020)Kazemi, Goel, Jain, Kobyzev, Sethi, Forsyth, and
  Poupart}]{kazemi20jmlr}
Kazemi, S.~M.; Goel, R.; Jain, K.; Kobyzev, I.; Sethi, A.; Forsyth, P.; and
  Poupart, P. 2020.
\newblock Representation learning for dynamic graphs: A survey.
\newblock \emph{Journal of Machine Learning Research} 21(70): 1--73.

\bibitem[{Ke, He, and Liu(2020)}]{ke2020rethinking}
Ke, G.; He, D.; and Liu, T.-Y. 2020.
\newblock Rethinking positional encoding in language pre-training.
\newblock \emph{arXiv preprint arXiv:2006.15595} .

\bibitem[{Keskar et~al.(2019)Keskar, McCann, Varshney, Xiong, and
  Socher}]{keskar2019ctrl}
Keskar, N.~S.; McCann, B.; Varshney, L.~R.; Xiong, C.; and Socher, R. 2019.
\newblock {CTRL}: A conditional Transformer language model for controllable
  generation.
\newblock \emph{arXiv preprint arXiv:1909.05858} .

\bibitem[{Kivel\"{a} et~al.(2014)Kivel\"{a}, Arenas, Barthelemy, Gleeson,
  Moreno, and Porter}]{kivela14jcn}
Kivel\"{a}, M.; Arenas, A.; Barthelemy, M.; Gleeson, J.~P.; Moreno, Y.; and
  Porter, M.~A. 2014.
\newblock {Multilayer networks}.
\newblock \emph{Journal of Complex Networks} 2(3): 203--271.

\bibitem[{Lerch et~al.(2019)Lerch, Arthur, Pati, and Gururani}]{alex2019music}
Lerch, A.; Arthur, C.; Pati, A.; and Gururani, S. 2019.
\newblock Music performance analysis: A survey.
\newblock In \emph{Proc. Int. Soc. Music Information Retrieval Conf.}

\bibitem[{Oore et~al.(2018)Oore, Simon, Dieleman, Eck, and
  Simonyan}]{oore2018time}
Oore, S.; Simon, I.; Dieleman, S.; Eck, D.; and Simonyan, K. 2018.
\newblock This time with feeling: Learning expressive musical performance.
\newblock \emph{Neural Computing and Applications} .

\bibitem[{Patel(2003)}]{patel03nature}
Patel, A.~D. 2003.
\newblock Language, music, syntax and the brain.
\newblock \emph{Nature Neuroscience} 6: 674--681.

\bibitem[{Payne(2019)}]{payne2019musenet}
Payne, C.~M. 2019.
\newblock {MuseNet}.
\newblock \emph{OpenAI Blog} .

\bibitem[{Rae et~al.(2020)Rae, Potapenko, Jayakumar, Hillier, and
  Lillicrap}]{rae20iclr}
Rae, J.~W.; Potapenko, A.; Jayakumar, S.~M.; Hillier, C.; and Lillicrap, T.~P.
  2020.
\newblock {Compressive Transformers} for long-range sequence modelling.
\newblock In \emph{Proc. Int. Conf. Learning Representations}.

\bibitem[{Raffel et~al.(2020)Raffel, Shazeer, Roberts, Lee, Narang, Matena,
  Zhou, Li, and Liu}]{raffel20jmlr_t5}
Raffel, C.; Shazeer, N.; Roberts, A.; Lee, K.; Narang, S.; Matena, M.; Zhou,
  Y.; Li, W.; and Liu, P.~J. 2020.
\newblock Exploring the limits of transfer learning with a unified text-to-text
  Transformer.
\newblock \emph{Journal of Machine Learning Research} 21(140): 1--67.

\bibitem[{Rossi et~al.(2020)Rossi, Chamberlain, Frasca, Eynard, Monti, and
  Bronstein}]{rossi20arxiv}
Rossi, E.; Chamberlain, B.; Frasca, F.; Eynard, D.; Monti, F.; and Bronstein,
  M. 2020.
\newblock Temporal graph networks for deep learning on dynamic graphs.
\newblock \emph{arXiv preprint arXiv:2006.10637} .

\bibitem[{Uitdenbogerd and Zobel(1999)}]{uitdenbogerd99mm}
Uitdenbogerd, A.; and Zobel, J. 1999.
\newblock Melodic matching techniques for large music databases.
\newblock In \emph{Proc. ACM Multimedia}, 57--66.

\bibitem[{Vaswani et~al.(2017)Vaswani, Shazeer, Parmar, Uszkoreit, Jones,
  Gomez, Kaiser, and Polosukhin}]{vaswani2017attention}
Vaswani, A.; Shazeer, N.; Parmar, N.; Uszkoreit, J.; Jones, L.; Gomez, A.~N.;
  Kaiser, {\L}.; and Polosukhin, I. 2017.
\newblock Attention is all you need.
\newblock In \emph{Proc. Advances in Neural Information Processing Systems},
  5998--6008.

\bibitem[{Wu and Yang(2020)}]{jazzTransformer20ismir}
Wu, S.-L.; and Yang, Y.-H. 2020.
\newblock The Jazz Transformer on the front line: Exploring the shortcomings of
  AI-composed music through quantitative measures.
\newblock In \emph{Proc. Int. Soc. Music Information Retrieval Conf.}

\bibitem[{Yang and Lerch(2020)}]{yang20evaluation}
Yang, L.-C.; and Lerch, A. 2020.
\newblock On the evaluation of generative models in music.
\newblock \emph{Neural Computing and Applications} 32: 4773--4784.

\end{thebibliography}

\end{document}